\documentclass[letter]{aa} 

\usepackage{graphicx}

\usepackage[varg]{txfonts}

\begin{document}

\title{Topological model of the anemone microflares \\
       in the solar chromosphere}

\titlerunning{Topological model of anemone microflares}

\author{
Yu.~V.~Dumin\inst{1,2}\fnmsep
  \thanks{\email{dumin@sai.msu.ru,dumin@yahoo.com}}
\and
B.~V.~Somov\inst{1}\fnmsep
  \thanks{\email{somov@sai.msu.ru,somov-boris@mail.ru}}
}

\institute{Sternberg Astronomical Institute (GAISh) of
Lomonosov Moscow State University,
Universitetskii prosp.\ 13, 119234, Moscow, Russia
  \and
Space Research Institute (IKI) of Russian Academy of Sciences,
Profsoyuznaya str.\ 84/32, 117997, Moscow, Russia}

\date{Received November 14, 2018; revised January 8, 2019
      and January 29, 2019; accepted February 14, 2019}

\abstract
{The chromospheric anemone microflares, which were discovered by Hinode
satellite about a decade ago, are specific transient phenomena starting
from a few luminous ribbons on the chromospheric surface and followed by
an eruption upward.
While the eruptive stage was studied in sufficient detail, a quantitative
theory of formation of the initial multi-ribbon structure remains
undeveloped until now.}
{We construct a sufficiently simple but general model of
the magnetic field sources that is able to reproduce all the observed
types of luminous ribbons by varying only a single parameter.}
{As a working tool, we employed the Gorbachev--Kel'ner--Somov--Shvarts (GKSS)
model of the magnetic field, which was originally suggested about three
decades ago to explain fast ignition of the magnetic reconnection
over considerable spatial scales by tiny displacements of the magnetic
sources.
Quite unexpectedly, this model turns out to be efficient for
the description of generic multi-ribbon structures in the anemone flares
as well.}
{As follows from our numerical simulation, displacement of a single
magnetic source (sunspot) with respect to three other sources results in
a complex transformation from three to four ribbons and then again to
three ribbons, but with an absolutely different arrangement.
Such structures closely resemble the observed patterns of emission in
the anemone microflares.}
{}


\maketitle

\section{Introduction}

One of the most interesting findings by the Hinode satellite
\citep{Kosugi_07,Tsuneta_08}, which was obtained soon after beginning
of its operation in 2006--2007, were the so-called anemone microflares
observed in the chromospheric \ion{Ca}{ii} line \citep{Shibata_07}.
Their development begins with three (or, less frequently, four) diverging
ribbons in the horizontal plane and then, at the second stage, the entire
structure erupts upward.
Because of their similarity with sea anemones, such phenomena were called
the anemone microflares.%
\footnote{At the same time, the term ``anemone'' was used also
to denote some types of active regions and the corresponding large
solar flares \citep{Asai_09}, which are unrelated to the microflares
discussed in the present paper.}

While the eruptive stage was studied in sufficient detail in a number of
subsequent publications \citep{Nishizuka_11,Singh_11,Singh_12}, much less
attention was paid to the formation of the initial multi-ribbon geometry.
In fact, it was interpreted until now only in terms of the qualitative pictures
of splitting magnetic tubes, such as in Fig.~3(D, E) by \citet{Shibata_07}.
In principle, it is not so difficult to invent some configuration of
the magnetic sources that produces the required splitting of the
magnetic fluxes.
However, an interesting question arises: Is it possible to suggest
a sufficiently simple but universal magnetic field model (e.g., involving
a single free parameter) that would be able to describe all the observed
types of anemone microflares?

Analysis of the various options that we performed recently has shown that
a reasonable choice might be based on the model by \citet{Gorbachev_88a},
which we refer to as GKSS.
This model was originally suggested as a mechanism for quick initiation
of the magnetic reconnection in a considerable volume of space under a
very small displacement of the magnetic sources, which was  called
the ``topological trigger'' of solar flares.%
\footnote{
Some more sophisticated topological models of the solar magnetic fields were
suggested, for example, by \citet{Inverarity_99} and \citet{Brown_01};
their review was given by \citet{Longcope_05}.
The applications of topological methods to various solar phenomena were
reviewed by \citet{Janvier_17}.}
Surprisingly, it turns out that the same model can also serve for
yet another purpose, i.e., the description of the multi-ribbon structures in
the anemone microflares.

\section{Theoretical model}

The GKSS model of magnetic fields was originally derived by exploiting the
quite sophisticated methods of differential geometry and algebraic topology,
whose description is far beyond the scope of the present letter;
for mathematical details, see the above-mentioned paper
by~\citet{Gorbachev_88a} and the later discussion
by~\citet{Brown_99} and \citet[][Sec.~4.2]{Somov_13}.
However, for the sake of completeness, we shall briefly summarize
some basic facts.

\begin{figure}
\centering
\includegraphics[width=0.9\hsize]{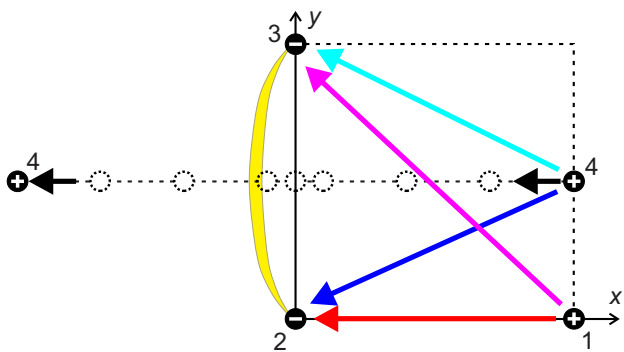}
\caption{
Arrangement of the positive and negative magnetic sources and
the corresponding semilunar region of topological instability (yellow)
in the GKSS model.
Four color arrows (red, magenta, blue, and cyan) designate four
different types of topological connectivity of the magnetic field
lines between these sources.
Various positions of the fourth magnetic source,
$ x_4 = 1 $, 0.7, 0.4, 0.1, 0.0, --0.1, --0.4, --0.7, and --1.0,
employed later, are shown by the black dotted circles.}
\label{Fig1}
\end{figure}

Let us consider the four point-like magnetic sources (idealized sunspots)
with equal magnitudes and opposite polarities (two positive and two
negative) in the plane $ z\,{=}\,0 $.
Three of these are localized in the vertices of an equilateral rectangular
triangle%
\footnote{
A cathetus of this triangle is used as a dimensionless unit of
length in the subsequent calculations.
}, as shown in Fig.~\ref{Fig1}, while the fourth source can take
different positions along the line $ y\,{=}\,0.5 $.

In the potential approximation%
\footnote{
From the physical point of view, the potential approximation implies that
the magnetic field in the volume under consideration is produced mostly by
the sources at its boundaries, while contribution by the bulk electric
currents is negligible.
Effects of the electric currents on global topology of the magnetic field
lines were analyzed, e.g., by \citet{Brown_00}.
}, the magnetic field formed by the above-mentioned sources can be
evidently written as
\begin{equation}
{\bf B} ({\bf r}) = \sum\limits_{i} e_i \, \frac{ {\bf r} - {\bf r}_i }%
                    { | {\bf r} - {\bf r}_i |^3 } \: ,
\label{eq:field_def}
\end{equation}
where $ e_i $~are the magnitudes of the sources and
$ {\bf r}_i $~are their positions.
In this work, $ e_1 = e_4 = 1 $ and $ e_2 = e_3 = -1 $.

\begin{figure}
\centering
\includegraphics[width=\hsize]{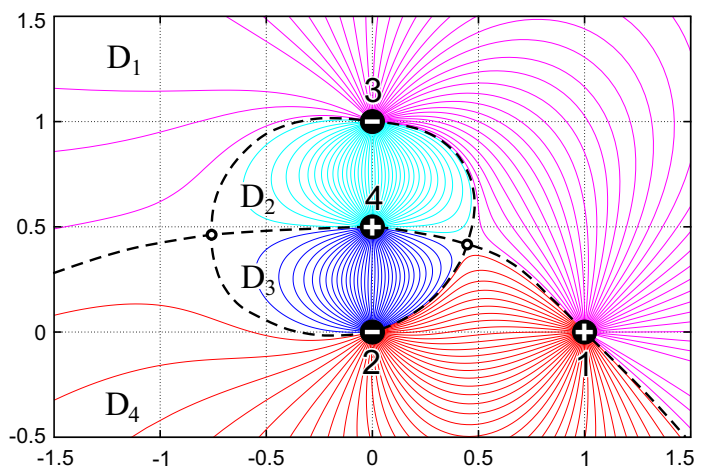}
\caption{
Base of the two-dome structure at $ z\,{=}\,0 $
(i.e., the ``two-oval structure'' drawn with dashed lines)
for the particular arrangement of the fourth source, $ x_4\,{=}\,0 $.
Magnetic field lines with different connectivity are shown in red, magenta,
blue, and cyan; these colors correspond to those in Fig.~\ref{Fig1}.
The first (small) oval is separated in the domains D$_2$ and D$_3$
by the large oval.
The domains D$_1$ and D$_4$ are outside the small oval and separated by
the large oval, where D$_1$~is the outer region.
Small circles at the intersection of the oval boundaries are
the null points of the magnetic field, which are the footpoints
of the separator.}
\label{Fig2}
\end{figure}

In general, the magnetic field in the entire half-space above the plane
of sources is given by the so-called ``two-dome configuration''.
Namely, four spatial subregions resulting from the two intersecting domes
(separatrices) correspond to the four types of connectivity of the magnetic
field lines, connecting the positive and negative sources.
As an example, Fig.~\ref{Fig2} shows a base of the two-dome structure
(i.e., its cross-section by the plane $ z\,{=}\,0 $).
The separatrices intersect each other in 3D space along the arc called
the separator, which originates from the null points in $ z\,{=}\,0 $
plane.

Therefore, the magnetic field lines connecting the fourth and second sources
fill the intersection of two domes, D$_3$; and the lines connecting
the fourth and third sources fill the small dome with the excluded
intersection region, D$_2$.
The lines connecting the first and second sources are within the region~D$_4$,
located in the large dome and outside the small dome.
Finally, the lines connecting the first and third sources are within
the outer space~D$_1$.

As was found by \citet{Gorbachev_88a}, when the fourth source is located in
the crescent region of ``topological instability'' (shown in yellow in
Fig.~\ref{Fig1}), the entire two-dome structure experiences a complex
transformation (a kind of flip), resulting in the appearance of
the additional (bifurcated) null point at the separator well above
the plane of sources; and its position considerably changes under a tiny
displacement of the fourth source \citep[for recent observational evidence
of this effect, resulting in the fast ignition of magnetic reconnection,
see][]{Dumin_17}.

\begin{figure}
\centering
\includegraphics[width=0.98\hsize]{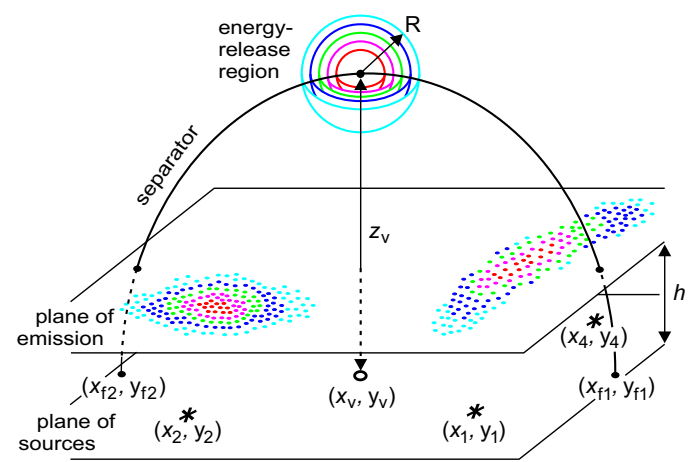}
\caption{Sketch of projections of the energy-release regions
with various radii~$ R $ (indicated by red, magenta, green, blue,
and cyan) along the magnetic field lines onto the plane
located at height~$ h $ above the field sources.
The sources are denoted by asterisks and the vertical projection of
the center of the energy-release region is denoted by a small thick circle.}
\label{Fig3}
\end{figure}

Yet another interesting phenomenon, immediately related to the interpretation
of anemone structures, can be found if we assume the existence of
an energy-release region at the top of the separator.
So, the energetic particles propagate downward along the magnetic
field lines and, losing their energy, produce the emission spots at
some height~$ h $; see Fig.~\ref{Fig3}.
This is actually the same idea that was used a long time ago by
\citet{Gorbachev_88b} for the interpretation of a two-ribbon structure in
large solar flares.
However, when the magnetic sources are arranged in the GKSS configuration,
the pattern of splitting of their fluxes and the respective emission spots
should be much more complex in the vicinity of the topologically unstable
region. Therefore, we can expect the appearance of multi-ribbon structures.
As shown in the next section, this hypothesis is well confirmed by
rigorous numerical calculations.

\section{Results of calculations}

\begin{figure}
\centering
\includegraphics[width=0.98\hsize]{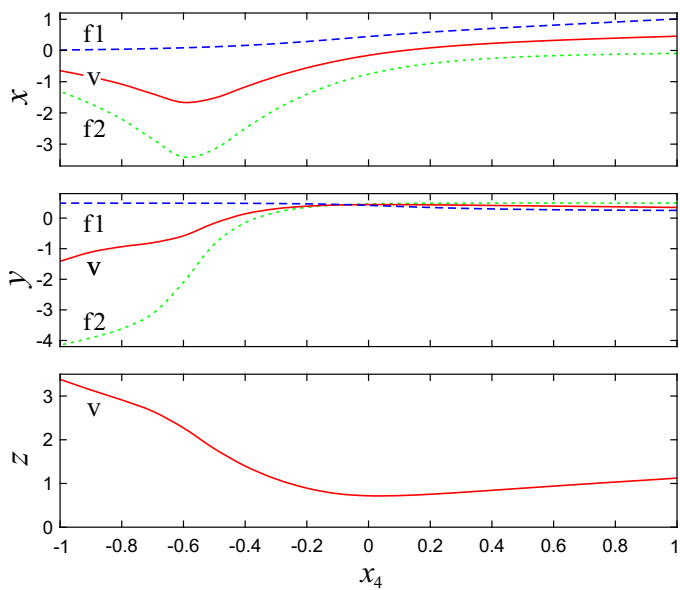}
\caption{
Coordinates $ x $, $ y $, and $ z $ (top, middle, and bottom panels,
respectively) of the first footpoint (dashed blue curves, denoted
by `f1'), second footpoint (dotted green curves, denoted by `f2'),
and vertex of the separator (solid red curve, denoted by `v') as
functions of the position of the fourth magnetic source~$ x_4 $.}
\label{Fig4}
\end{figure}

The basic domain of computation is presented in Fig.~\ref{Fig3}:
The point-like magnetic sources with coordinates $ (x_1, y_1) $,
$ (x_2, y_2) $, $ (x_3, y_3) $, and $ (x_4, y_4) $ are located in
the plane $ z\,{=}\,0 $; these sources can be physically interpreted as
sunspots in the solar photosphere.
The separator, where two ``magnetic domes'' (separatrices) intersect
each other, has footpoints $ (x_{\rm f1}, y_{\rm f1}) $ and
$ (x_{\rm f2}, y_{\rm f2}) $ in the same plane, and its vertex is
in the point $ (x_{\rm v}, y_{\rm v}, z_{\rm v}) $ above this plane.
Their numerical values at various positions of the fourth source~$ x_4$ are
listed in Table~\ref{Table1};
the algorithm employed for their computation is outlined in
Appendix~\ref{sec:Separator}.
In addition, all these characteristic points of the separator as functions
of~$ x_4 $ are pictorially represented in Fig.~\ref{Fig4}.

Then, we consider a set of the energy-release regions in the form
of balls with various radii~$ R $, whose centers are in the vertex of
the separator $ (x_{\rm v}, y_{\rm v}, z_{\rm v}) $; they are indicated
by different colors in Fig.~\ref{Fig3}.
Next, we take the initial points uniformly distributed over the
surface of the corresponding ball (with mutual separation,
for example,~$ 15^\circ $) and integrate the standard equations for
the magnetic field lines,
\begin{equation}
\frac{{\rm d}{\bf r}(l)}{{\rm d}l} =
  \frac{{\bf B}({\bf r}(l))}{|{\bf B}({\bf r}(l))|} \, ,
\label{eq:field_line}
\end{equation}
either in forward or backward direction (i.e., increasing or decreasing~$ l $)
until they intersect the plane located below at height~$ h $.
From the physical point of view, this means that charged particles are
accelerated in the energy-release region, propagate along the field lines,
and finally are decelerated and lose their energy in the denser layer of
the solar atmosphere at $ z\,{=}\,h $.
Hence, the corresponding plane is called the plane of emission.
Projections of the energy-release regions onto this plane are shown by
the dotted areas of the respective colors.
They have a granular structure in Fig.~\ref{Fig3} only because we traced
a finite number of magnetic field lines.

\begin{figure*}
\centering
\includegraphics[width=0.98\hsize]{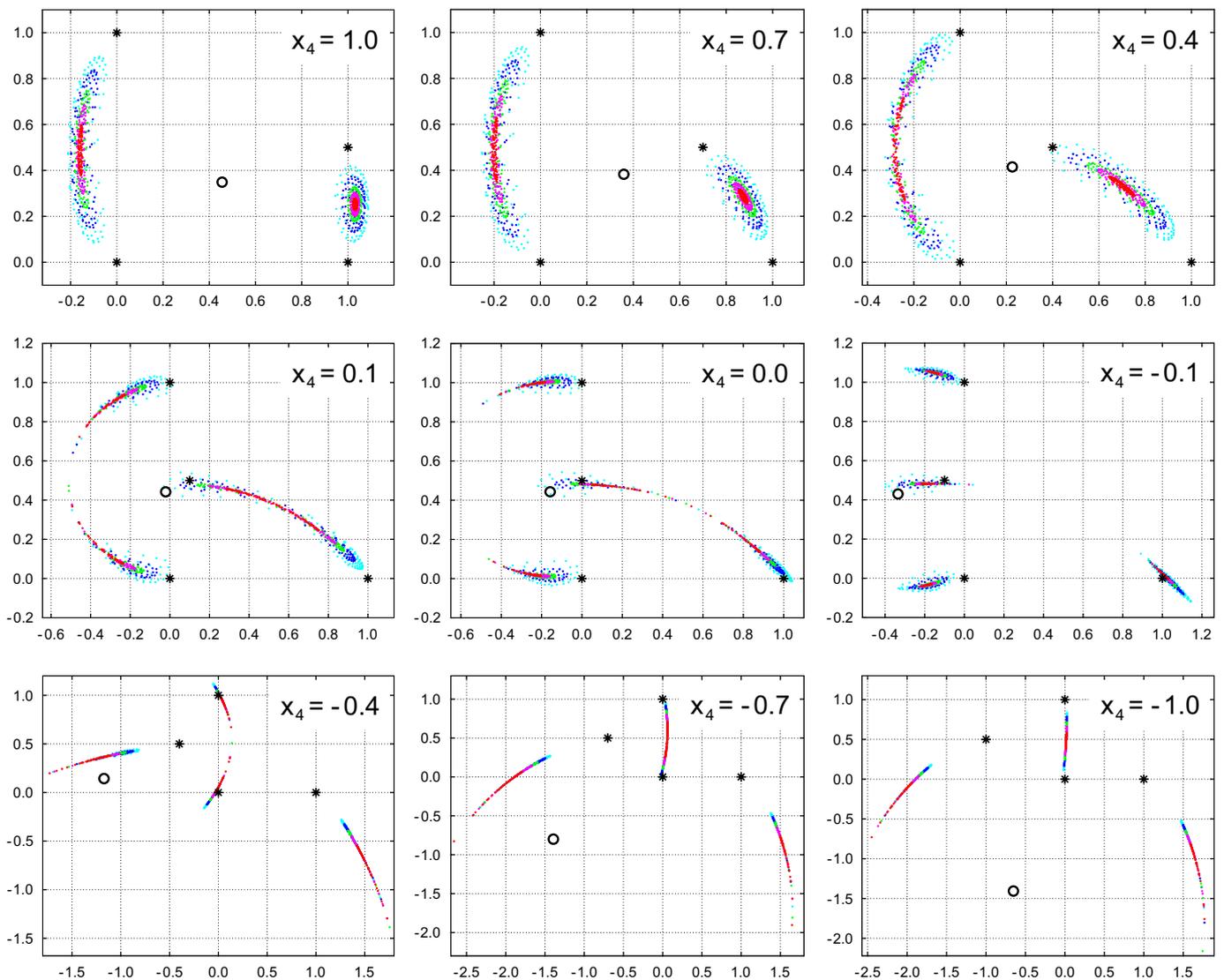}
\caption{Calculated projections of the energy-release region
(taken at the top of the separator) along the magnetic field lines
onto the horizontal plane (located at the height $ h = 0.3 $ above
the plane of magnetic sources) at various positions of the fourth
source~$ x_4 $. The red, magenta, green, blue, and cyan points
correspond to the various radii of the energy-release region
$ R\,{=}\,0.05 $, 0.10, 0.15, 0.25, and 0.35, respectively.
Positions of the magnetic field sources are denoted by the asterisks,
and vertical projections of the energy-release regions are
shown by the small circles.}
\label{Fig5}
\end{figure*}

The numerical calculations of the above-mentioned projections were performed
for different positions of the fourth magnetic source $ x\,{\in}\,[-1, 1] $
along the line $ y\,{=}\,0.5 $.
Along with that, we studied how the pattern of projections depend on
the parameters~$ R $ and $ h $.
A particular example for
$ x_4\,{=}\,1.0 $, 0.7, 0.4, 0.1, 0.0, $-0.1$, $-0.4$, $-0.7$, $-1.0$,\:
$ R\,{=}\,0.05 $, 0.10, 0.15, 0.25, 0.35,\: and $ h\,{=}\,0.3 $
(everything in dimensionless units, normalized to the separation
between the magnetic sources) is presented in Fig.~\ref{Fig5}.

\section{Discussion and conclusions}

As is seen in Fig.~\ref{Fig5}, the spots of emission in the $ h $-plane
(which can be interpreted as chromosphere, located above the photospheric
sources) experience a drastic transformation when the fourth magnetic source
shifts to the left, as depicted in Fig.~\ref{Fig1}:
\begin{itemize}
\item
Initially, at $ x_4\,{=}\,1.0 $, 0.7, and 0.4, the situation resembles
the classical two-ribbon flare, as was found earlier by
\citet{Gorbachev_88b}.
\item
Next, near the region of topological instability ($ x_4\,{=}\,0.1 $ and 0.0),
these ribbons become distorted and subdivided into three parts.
\item
Then, immediately in the region of topological instability
($ x_4\,{=}\,{-}0.1 $), we get a four-ribbon structure.
\item
At last, when the fourth source moves further to the left
($ x_4\,{=}\,{-}0.4 $, $-0.7$, and $-1.0$), the four ribbons merge again into
three ribbons, but with an absolutely different spatial arrangement.
\end{itemize}
The above-listed configurations closely resemble chromospheric anemone
flares, possessing usually three and, less frequently, four ribbons,
as seen in the numerous movies available in the Hinode archive.%
\footnote{
\tiny \url{https://hinode.nao.ac.jp/en/for-researchers/qlmovies/top.html}}
To avoid misunderstanding, let us emphasize that various values of~$ x_4 $
correspond to a statistical sample of various microflares.
They should not be interpreted as a real physical motion of the fourth source
in the course of the flare development, since the lifetime of such flares
is very short.

As regards the dependence of the emission ribbons on the radii~$ R $, we can
see a quite complicated behavior.
In the two- and four-ribbon configurations, a core of the energy-release
region (i.e., the balls with smaller~$ R $) is projected into the centers
of the ribbons and the outer part (with larger~$ R $) into their periphery
(see panels with $ x_4\,{=}\,1.0 $, 0.7, 0.4, and $-0.1$ in Fig.~\ref{Fig5}).
However, in the three-ribbon configurations, both on the right- and
left-hand side of the topological instability region, the situation
is different. One ribbon looks as described above while in two other
ribbons the inner and outer parts of the energy-release region tend
to project into the opposite sides of the strips (see panels with
$ x_4\,{=}\,0.1 $, $-0.7$, and $-1.0$).

At last, as follows from our computations of the emission ribbons at
various heights~$ h $ (not represented in detail in this Letter),
the corresponding patterns remain qualitatively the same such that only
the degree of expression of the ribbons (i.e., the sharpness and
separation between them) varies.
In fact, the best expression takes place just at the height about
$ h\,{=}\,0.3 $, which is presented in Fig.~\ref{Fig5}.

In principle, it might also be interesting to study the dependence of
the emission spots on the position of the energy-release region along
the separator.
However, in this paper we studied only the case of very small magnetic
loops, corresponding to the microflares. In fact, a height of the arc was
only two to three times greater than the height of the plane of emission above
the plane of sources, and the size of the energy-release region was
of the same order of magnitude.
Therefore, in these conditions, it was impossible to analyze any substantial
shifts of the energy-release region along the separator.
Nevertheless, some hints concerning this issue can be drawn from the behavior
of the emission spots as function of the radii of the energy-release
spheres~$ R $, which are plotted by various colors in Figs.~\ref{Fig3}
and~\ref{Fig5}.

We also checked the patterns of emission obtained under the similar
assumptions in a few other topological models of solar magnetic fields
(i.e., involving the bifurcations of null points).
In these models, however, we did not find the behavior of the ribbons
as rich as in the GKSS model.
This is most probably because only the latter model exhibits the genuine
topological instability, i.e., when a small rearrangement of the magnetic
sources results in the dramatic reconstruction of the magnetic field lines
in the entire space.

In summary, the four-ribbon configurations exist only in a quite small
semilunar region of topological instability (where the bifurcation takes
place); at all other locations of the fourth magnetic source, a three-
(or even two-)ribbon emission appears.
Hence, we should expect that fraction of the four-ribbon flares will be
sufficiently small because of the relatively small area of the region
of topological instability.
This prediction agrees very well with the observations.
Most of the anemone flares detected by Hinode satellite possess three ribbons;
the four-ribbon flares also exist but they are formed seldomly.
This is why we believe that our topological model, although not universal,
provides a reasonable interpretation of the relevant phenomena by varying
only a single parameter.

\begin{acknowledgements}

YVD is grateful to P.M.~Akhmet'ev and E.V.~Zhuzhoma for consultations on
the topological issues, as well as to A.V.~Getling, A.V.~Oreshina, and
I.V.~Oreshina for help in processing the data on magnetic fields.

BVS was supported by the Russian Foundation for Basic Research, grant
no.~16-02-00585.

\end{acknowledgements}

\newpage

\begin{appendix}

\section{Basic parameters of the separator in the two-dome structure}
\label{sec:Separator}

\begin{table*}
\caption[]{Computed coordinates of the separator's footpoints
$ (x_{\rm f1}, y_{\rm f1}, 0) $ and $ (x_{\rm f2}, y_{\rm f2}, 0) $
and of its vertex $ (x_{\rm v}, y_{\rm v}, z_{\rm v}) $,
where the energy-release region is located,
at various positions~$ x_4 $ of the fourth magnetic field source.}
\label{Table1}
\centering
\begin{tabular}{rrrrrrrrrrr}
\hline \\[-9pt]
\multicolumn{1}{c}{source} &&
\multicolumn{2}{c}{1st footpoint} &&
\multicolumn{2}{c}{2nd footpoint} &&
\multicolumn{3}{c}{vertex}
\\
\multicolumn{1}{c}{$ x_4 $} &&
\multicolumn{1}{c}{$ x_{\rm f1} $} &
\multicolumn{1}{c}{$ y_{\rm f1} $} &&
\multicolumn{1}{c}{$ x_{\rm f2} $} &
\multicolumn{1}{c}{$ y_{\rm f2} $} &&
\multicolumn{1}{c}{$ x_{\rm v} $}  &
\multicolumn{1}{c}{$ y_{\rm v} $}  &
\multicolumn{1}{c}{$ z_{\rm v} $}
\\
\hline \\[-9pt]
  1.0000 &&   1.0110 &   0.2506 && --0.0965 &   0.4904 &&
   0.457 &   0.347 &   1.123 \\
  0.9000 &&   0.9615 &   0.2542 && --0.1074 &   0.4905 &&
   0.427 &   0.359 &   1.077 \\
  0.8000 &&   0.9119 &   0.2587 && --0.1216 &   0.4905 &&
   0.395 &   0.370 &   1.033 \\
  0.7000 &&   0.8620 &   0.2648 && --0.1406 &   0.4905 &&
   0.361 &   0.381 &   0.987 \\
  0.6000 &&   0.8115 &   0.2730 && --0.1664 &   0.4905 &&
   0.323 &   0.391 &   0.940 \\
  0.5000 &&   0.7597 &   0.2843 && --0.2018 &   0.4903 &&
   0.279 &   0.402 &   0.892 \\
  0.4000 &&   0.7059 &   0.2997 && --0.2513 &   0.4897 &&
   0.227 &   0.413 &   0.844 \\
  0.3000 &&   0.6493 &   0.3204 && --0.3210 &   0.4883 &&
   0.164 &   0.424 &   0.797 \\
  0.2000 &&   0.5885 &   0.3475 && --0.4197 &   0.4851 &&
   0.084 &   0.433 &   0.754 \\
  0.1000 &&   0.5216 &   0.3807 && --0.5600 &   0.4781 &&
 --0.019 &   0.440 &   0.723 \\
  0.0000 &&   0.4467 &   0.4160 && --0.7576 &   0.4623 &&
 --0.156 &   0.441 &   0.718 \\
--0.1000 &&   0.3658 &   0.4459 && --1.0312 &   0.4269 &&
 --0.333 &   0.428 &   0.766 \\
--0.2000 &&   0.2865 &   0.4654 && --1.4005 &   0.3497 &&
 --0.557 &   0.389 &   0.893 \\
--0.3000 &&   0.2173 &   0.4762 && --1.8854 &   0.1870 &&
 --0.834 &   0.305 &   1.102 \\
--0.4000 &&   0.1615 &   0.4818 && --2.4968 & --0.1509 &&
 --1.168 &   0.139 &   1.399 \\
--0.5000 &&   0.1181 &   0.4849 && --3.1644 & --0.8531 &&
 --1.523 & --0.167 &   1.797 \\
--0.6000 &&   0.0850 &   0.4866 && --3.4129 & --2.0948 &&
 --1.664 & --0.580 &   2.271 \\
--0.7000 &&   0.0598 &   0.4876 && --2.8354 & --3.1245 &&
 --1.388 & --0.805 &   2.653 \\
--0.8000 &&   0.0406 &   0.4882 && --2.2016 & --3.6193 &&
 --1.081 & --0.935 &   2.911 \\
--0.9000 &&   0.0259 &   0.4886 && --1.7097 & --3.9167 &&
 --0.842 & --1.112 &   3.139 \\
--1.0000 &&   0.0144 &   0.4888 && --1.3089 & --4.1499 &&
 --0.647 & --1.411 &   3.383 \\
\hline
\end{tabular}
\end{table*}

In general, computing a spatial configuration of the separator is
a non-trivial task: the separator is an inherently unstable structure,
since any field line originating in its neighborhood quickly deviates
outward.
So, the straightforward numerical integration is meaningless, and a more
elaborate type of algorithm should be applied.
In the present work, we used the following procedure:
\begin{enumerate}
\item
Since footpoints of the separator, $ (x_{\rm f1}, y_{\rm f1}) $ and
$ (x_{\rm f2}, y_{\rm f2}) $, are actually the null points of
the magnetic field in $ z\,{=}\,0 $ plane, we calculated the distribution
of~$ B^2 $ in this plane, sought for its minima, and finally checked
that~$ B^2\,{=}\,0 $ in these minima (and, therefore, $ {\bf B}\,{=}\,0 $).
The analysis of the scalar function~$ B^2(x,y) $ was employed only because,
from the computational point of view, it is much more simple and robust than
the analysis of the vector field~$ {\bf B}(x,y) $.
\item
To find a position of the separator above the plane $ z\,{=}\,0 $, we analyzed
the following ``global'' behavior of the magnetic field lines:
\begin{enumerate}
\item
First of all, we fixed a vertical plane intersecting the line connecting
the opposite footpoints of the separator $ (x_{\rm f1}, y_{\rm f1}, 0) $ and
$ (x_{\rm f2}, y_{\rm f2}, 0) $;
to get a better accuracy, it is desirable that this plane be approximately
perpendicular to the line.
\item
Then, we generated a mesh of points in the above-mentioned plane; and
each of these points was used as the initial condition for integration of
the magnetic field line equation~(\ref{eq:field_line}) in the forward and
backward directions until the corresponding field line entered one of
the magnetic sources $ e_1 $, $ e_2 $, $ e_3 $, or $ e_4 $.
Thereby, the sketch of topological connectivity, composed of four subregions
whose field lines terminate at four different sources, is obtained.
These subregions touch each other in some spot, which is just the place
where separator intersects the specified plane.
\item
Next, we fixed a smaller rectangle covering the above-mentioned spot,
generated a finer mesh of initial points, and performed the same procedure
as in the previous item.
\item
Repeating these steps a few times, we can localize the spot of intersection
of the separator with the specified plane as accurately as desired.
\item
If necessary, the points of intersection with a set of planes can be
interpolated between each other to get the intermediate positions of
the separator.
\end{enumerate}
\end{enumerate}

The most important parameters of the resulting separators---the coordinates
of their footpoints and vertices---at various positions of the
fourth source~$ (x_4, 0.5) $ are listed in Table~\ref{Table1} and drawn in
Fig.~\ref{Fig4}.

\end{appendix}

\end{document}